\documentclass[10pt,letterpaper]{article}
\usepackage[top=0.85in,left=2.75in,footskip=0.75in]{geometry}

\usepackage{changepage}

\usepackage[utf8]{inputenc}

\usepackage{textcomp,marvosym}

\usepackage{fixltx2e}

\usepackage{amsmath,amssymb}

\usepackage{cite}

\usepackage{nameref,hyperref}

\usepackage[right]{lineno}

\usepackage{microtype}
\DisableLigatures[f]{encoding = *, family = * }

\usepackage{rotating}

\raggedright
\usepackage[aboveskip=1pt,labelfont=bf,labelsep=period,justification=raggedright,singlelinecheck=off]{caption}

\makeatletter
\renewcommand{\@biblabel}[1]{\quad#1.}
\makeatother

\date{}

\usepackage{lastpage,fancyhdr,graphicx}
\begin{document}
\vspace*{0.35in}

\begin{flushleft}
{\Large
\textbf\newline{Outcome prediction in mathematical models of immune response to infection}
}
\newline
\\
Manuel Mai\textsuperscript{1,*},
Kun Wang\textsuperscript{2,3},
Greg Huber\textsuperscript{4},
Michael Kirby\textsuperscript{2,5},
Mark~D.~Shattuck\textsuperscript{6,3},
Corey S. O'Hern\textsuperscript{1,3,7,8}
\\
\bf{1} Department of Physics, Yale University, New Haven, CT, USA
\\
\bf{2} Department of Mathematics, Colorado State University, Fort Collins, CO, USA
\\
\bf{3} Department of Mechanical Engineering and Material Science, Yale University, New Haven, CT, USA
\\
\bf{4} Kavli Institute for Theoretical Physics, Kohn Hall, University of California, Santa Barbara, CA, USA
\\
\bf{5} Department of Computer Science, Colorado State University, Fort Collins, CO, USA
\\
\bf{6} Benjamin Levich Institute and Physics Department, The City College of New York, New York, NY, USA
\\
\bf{7} Department of Applied Physics, Yale University, New Haven, CT, USA
\\
\bf{8} Graduate Program in Computational Biology and Bioinformatics, Yale University, New Haven, CT, USA
\\

%
%





* E-mail: manuel.mai@yale.edu
\end{flushleft}
\section*{Abstract}
Clinicians need to predict patient outcomes with high accuracy as
early as possible after disease inception.  In this manuscript, we
show that patient-to-patient variability sets a fundamental limit on
outcome prediction accuracy for a general class of mathematical models
for the immune response to infection. However, accuracy can be
increased at the expense of delayed prognosis.  We investigate several
systems of ordinary differential equations (ODEs) that model the host
immune response to a pathogen load.  Advantages of systems of ODEs for
investigating the immune response to infection include the ability to
collect data on large numbers of `virtual patients', each with a given
set of model parameters, and obtain many time points during the course
of the infection. We implement patient-to-patient variability $v$ in
the ODE models by randomly selecting the model parameters from
Gaussian distributions with variance $v$ that are centered on
physiological values.  We use logistic regression with one-versus-all
classification to predict the discrete steady-state outcomes of the
system. We find that the prediction algorithm achieves near $100\%$
accuracy for $v=0$, and the accuracy decreases with increasing $v$ for
all ODE models studied.  The fact that multiple steady-state outcomes
can be obtained for a given initial condition, {\it i.e.} the basins
of attraction overlap in the space of initial conditions, limits the
prediction accuracy for $v>0$.  Increasing the elapsed time of the
variables used to train and test the classifier, increases the
prediction accuracy, while adding explicit external noise to the ODE
models decreases the prediction accuracy. Our results quantify the
competition between early prognosis and high prediction accuracy that
is frequently encountered by clinicians.


\section*{Introduction}
\label{intro}

The immune response to infection is a complex process that involves a
wide range of length scales from proteins to cells
\cite{koup1994,guidotti2001,janeway2002,klipp2009},
tissues\cite{masopust2001}, and organ systems\cite{pope2001}. Despite
enormous progress over the past 30 years in developing mathematical
models for the immune response to infectious disease such as
tuberculosis\cite{young2008},
HIV\cite{perelson1999,callaway2002,nelson2002,perelson2002}, and
influenza\cite{bocharov1994,belz2002}, these models still have not been able
to dramatically improve patient diagnosis, prognosis, and
treatment\cite{dowdy2006,Wodarz2002}.  Instead, vaccine and drug
development often relies on costly trial-and-error
methods\cite{serdobova2006}.  However, advances in gene sequencing
capabilities\cite{wetterstrand}, increasing speeds of computer
processors, and the ability to store enormous amounts of medical data
promise dramatic improvements in mathematical
approaches to predicting and controlling the response to infectious
disease~\cite{Day2006,vodovotz2008,vodovotz2009}.

One promising mathematical approach is to use machine learning methods
on large data sets to classify patients as healthy or sick, perform
early warning analyses for early detection of infection, or identify
the minimal set of genes responsible for a particular immune
response.\cite{wang2013,ohara2013} However, many questions are left
unanswered in such studies.  For example, how much and what kinds of
data are required to have confidence in the machine learning
predictions and what are the underlying biophysical mechanisms for the
relationships between variables that are identified by these
techniques?  Further, it is difficult to determine differences in
the immune response that arise from patient-to-patient variations compared
to slight differences in the initial conditions of each patient.

In this manuscript, we focus on sets of ordinary differential
equations (ODEs) as mathematical models for the immune response to
infection.  The advantages of ODEs are manifold: 1) Each `virtual
patient' can be considered as a set of parameters in the set of ODEs;
2) There is essentially no limit on the amount of data that can be
collected on each virtual patient; 3) The accuracy of machine learning
predictions can be explicitly tested as a function of the number of
time points and initial conditions for each patient and the number of
patients included in the training and testing sets; and 4) analysis of
the fixed points (or steady-state outcomes) and basins of attraction
of the ODEs can give biophysical insight into the immune response to
infection.
 
We will investigate several classes of ODE models for the immune
response to infection. First, we will describe a four-dimensional
model for the acute inflammatory response to a pathogen load that was
studied in detail in Ref.~\cite{reynolds2006}.  We will then consider
reduced versions of this model with fewer variables and parameters
obtained by slaving one or more of the original four variables, as
well as changes to form of the ODEs that alter the fixed point
structure and flows between them. For each model, a virtual patient is
defined by one set of parameters. Given an initial condition (values
of the variables at time $t=0$), the patient will evolve
deterministically to one of several possible discrete steady-state ($t
\rightarrow \infty$) outcomes, or fixed points.  Thus, for each
patient, we can determine the basins of attraction that map initial
conditions for all of the variables to steady-state outcomes by
numerically integrating the sets of ODEs.

\begin{figure}[!ht]
\begin{center}
\includegraphics[width=5.25in]{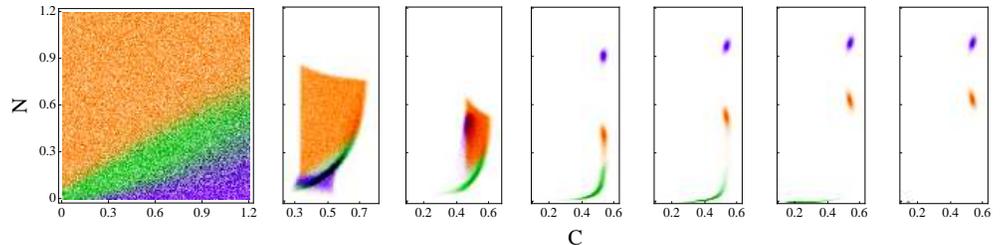}
\end{center}
\vspace{-0.2in}
\caption{ {\bf Time evolution of patient outcomes for a range of
neutrophil ($N$) and cortisol ($C$) initial conditions for Model $A$.}
(left) Patient outcomes in the long-time limit given $N$ and $C$
initial conditions for model (1) are shaded green, orange, and purple
for the steady-state outcomes of health, aseptic, and septic death,
respectively. The initial values of the pathogen load and damage are
$P_0 = 0.35$ and $D_0 = 0$, and patient variability is set to
$v=2\%$. The right six panels indicate how the systems in the leftmost
panel separate in the $N$ and $C$ plane as time increases, $t=10$,
$20$, $50$, $100$, $250$, and $500$, from left to right.}
\label{fi:schematic}
\end{figure}

We seek to determine the limits of the prediction accuracy of discrete
steady-state outcomes of ODEs as a function of patient variability
({\it i.e.} random fluctuations in parameter values) using machine
learning techniques.  In the limit of zero patient variability, our
simple classification algorithm (logistic regression) can achieve
nearly perfect prediction accuracy even when the classification occurs
on variables at short times. However, as the patient variability
increases, the basins of attraction for different patients yield
different outcomes for a given initial condition as shown in
Fig.~\ref{fi:schematic} for $2\%$ patient variability in model (1) for
the immune response to infection. (See Materials and Methods.)  The
fact that each set of initial conditions does not possess a unique
outcome places a fundamental limit on the predictability of patient
outcomes.  Thus, we find that the machine learning prediction accuracy
decreases with increasing patient variability.  In contrast, for a
given patient variability, the prediction accuracy increases with the
time used for classification as the systems converge to their
steady-state outcomes (Fig.~\ref{fi:schematic}).  We also show that at
short times our classification algorithm saturates the theoretical
limit for the prediction accuracy in the presence of patient
variation, and that the addition of external noise only worsens the
outcome prediction accuracy.

The manuscript is organized as follows. In the Materials and Methods
section, we introduce several ODE systems that have been used to model
the host immune response to infection~\cite{reynolds2006}, including
their parameter sets and discrete steady-state outcomes, and describe how we
implement patient variability in the ODE models.  In the Results
section, we emphasize our three main results that hold for all of the
ODE models we studied: 1) patient variability leads to overlap of the
basins of attraction for the steady-state outcomes, which limits the
outcome prediction accuracy, 2) the prediction accuracy increases with
the time used for classification because the basins of attraction
separate with increasing time, and 3) the addition of external
measurement noise further reduces the prediction accuracy.  In the
Discussion section, we point out the clinical implications of our work
and describe important future studies of the prediction accuracy for
ODE models with continuous outcomes. In the Supporting Information, we
discuss a generalized one-dimensional ODE against which we compare our
results and the numerical implementation of the algorithms used to
classify the steady-state outcomes of the ODEs.

\section*{Materials and Methods}

Our studies focus on several ODE models with varying complexity
for the immune response to pathogen load that were first
introduced in Ref.~\cite{reynolds2006}. These ODE models can have up
to four coupled variables that represent the concentration of pathogen
$P$, activated neutrophils $N$, inflammation (or damage) $D$, and
immuno-suppressor (cortisol) $C$.  The models include interactions
between these four quanties.  For example, the presence of pathogen
$P>0$ causes an immune response, where neutrophils are activated and
$N$ increases. Neutrophils kill pathogen, which decreases $P$, but
also cause inflammation (damage), which increases $D$. The cortisol
level $C$ increases when there is a high neutrophil level, which then
reduces the neutrophil level.  

Model (1) (Eqs.~\ref{eq:mod1P}-\ref{eq:mod1C}) includes all four
variables $P$, $N$, $D$, and $C$.  The right-hand side of $dP/dt $ is
a sum of three terms. The first term enables logistic growth of the
pathogen. In the absence of any other terms, any positive initial
$P_0$ will cause $P$ to grow logistically to the steady-state value
$P_\infty$. The second term mimics a local, non-specific response to
an infection. For small values of $P$, the decrease is proportional to
$P$. For larger values of $P$, the decrease caused by the second term
is constant. The third term models the decrease of $P$ due to
interactions with activated immune cells (neutrophils) $N$. Activated
neutrophils $N$ can directly decrease $P$. The anti-inflammatory
response, which is captured by the cortisol level $C$, mitigates this
effect leading to a decrease of $P$ proportional to
$N*P/(1+(C/C_\infty)^2)$.

Two terms determine the rate of change in neutrophils, $dN/dt$. The
first term accounts for the fact that neutrophils can be activated if a
resting neutrophil cell encounters a pathogen $P$ or an already
activated neutrophil $N$. Furthermore, tissue damage $D$ also triggers
the activation of neutrophils. The second term describes the death of
neutrophils $N$, with the decrease in $N$ proportional to the amount of
neutrophils present.

The rate of change in damage $dD/dt$ is also controlled by two
terms. The first term mimics positive feedback between $D$ and
$N$. Activated phagocytes cause collateral damage in the
tissue. Again, the effectiveness of $N$ is mitigated by the
anti-inflammatory response $1/(1+(C/C_\infty)^2)$. The saturation
function $f_s$ models the fact that the effect of $N$ on $D$ saturates
for large $N$. The second term, -$\mu_d D$, represents
repair of the tissue.

The anti-inflammatory response $C$ increases with the source
term $s_c$. In addition, there is a natural death rate $\mu_c$, which leads to
a positive steady-state value of $C$ in the absence of any immune
activation $N$ or damage $D$. However, even small amounts of damage and
neutrophils will up-regulate $C$. In the case of small $N + k_{cnd}D$, the
production of $C$ is proportional to $N + k_{cnd}D$, while for large
values of $N + k_{cnd}D$, changes in $C$ are proportional to $k_{cn}$. Again,
the effectiveness of $N$ is mitigated by $1/(1+(C/C_\infty)^2)$.

Model (1) has $21$ parameters: $k_{pm}$, $k_{mp}$, $s_m$, $\mu_m$,
$k_{pg}$, $P_{\infty}$, $k_{pn}$, $k_{np}$, $k_{nn}$, $s_{nr}$,
$\mu_{nr}$, $\mu_n$, $k_{nd}$, $k_{dn}$, $x_{dn}$, $\mu_d$,
$C_{\infty}$, $s_c$, $k_{cn}$, $k_{cnd}$, and $\mu_c$. Depending on
the values of these parameters, model (1) possesses different numbers
of fixed points with varying stabilities.  However, we will focus on a
specific parameter regime (given in Table~\ref{tab:parameters}) with
three stable fixed points, which correspond to the physiological
steady-state outcomes: health, septic death, and aseptic death.

\begin{table}[!ht]
\caption{
{\bf Parameter mean values for model (1)}}
\begin{tabular}{|c|c|}
\hline
Parameter $q$ & Mean Value $\mu_q$\\
\hline
$k_{pm}$& 0.6\\
\hline
$k_{mp}$& 0.01 \\
\hline
$s_{m}$& 0.005 \\
\hline
$\mu_{m}$& 0.002 \\
\hline
$k_{pg}$& 0.6 \\
\hline
$P_\infty$& 20.0 \\
\hline
$k_{pn}$& 1.8 \\
\hline
$k_{np}$& 0.1 \\
\hline
$k_{nn}$& 0.01 \\
\hline
$s_{nr}$& 0.08 \\
\hline
$\mu_{nr}$& 0.12 \\
\hline
$\mu_{n}$& 0.05 \\
\hline
$k_{nd}$& 0.02 \\
\hline
$k_{dn}$& 0.35 \\
\hline
$x_{dn}$& 0.06 \\
\hline
$\mu_{d}$& 0.02 \\
\hline
$C_{\infty}$& 0.28 \\
\hline
$s_{c}$& 0.0125\\
\hline
$k_{cn}$& 0.04 \\
\hline
$k_{cnd}$& 48.0 \\
\hline
$\mu_{c}$& 0.1 \\
\hline
\end{tabular}
\label{tab:parameters}
\end{table}

\paragraph{Model (1)}   
\begin{align}
\label{eq:mod1P}
\frac{dP}{dt}&= \,      k_{pg}P \left(1-\frac{P}{P_\infty}\right)-\frac{k_{pm} s_m P}{\mu_m+k_{mp} P}-k_{pn}f(N)P\\
\label{eq:mod1N}
\frac{dN}{dt}&=\,       \frac{s_{nr}R}{\mu_{nr}+R}-\mu_nN\\
\label{eq:mod1D}
\frac{dD}{dt}&=\,               k_{dn}f_s\left(f(N)\right)-\mu_dD\\
\label{eq:mod1C}
\frac{dC}{dt}&=\,       s_c + \frac{k_{cn}\,f(N+k_{cnd} D)}{1+f(N+k_{cnd}D)}    -\mu_c C,
\end{align}
\quad \quad where 
\begin{align*}
R&=f(k_{nn} N+k_{np}P+k_{nd}D),\\
f(V)&=V/(1+(C/C_\infty)^2),\\
f_s(V)&=V^6/(x^6_{nd}+V^6).
\end{align*}

Models (2)-(5) given below are simplified versions of model (1).  A
summary of the dimension, number of parameters, and number of stable fixed
points for each of the ODE models is shown in
Table~\ref{tab:modelsum}.  To obtain model (2) from (1), $C$ is set to
a constant ${\overline C}=0.23$ and the remaining terms define a
three-variable model with $P$, $N$, and $D$. For model (3), we set
$C={\overline C}$ and $D=0$, which gives a two-variable model for $P$
and $N$.  For model (4), we set $C={\overline C}$ and $P=0$ to obtain
a two-variable model for $N$ and $D$. In this model, the value of the
initial rise in $N$ can be thought of as the response to trauma.  For
model (5), we set $C={\overline C}=0.1$, $D=0$, and $N=0$, which gives a
one-dimensional model for $P$.  This model only treats the innate
immune response with no activated neutrophils.

\begin{table}[!ht]
\caption{
{\bf Summary of ODE models}}
\begin{tabular}{|r|c|c|c|c|c|}
\hline
Model & Dimension & Parameters  q& Stable Fixed Points & $k_{pg}$ & ${\overline C}$\\
\hline
1 & 4 & 21 & 3 & 0.6 & N/A\\
\hline
2 & 3 & 18 & 3 & 1.2 & 0.23\\
\hline
3 & 2 & 14 & 2 & 1.2 & 0.23\\
\hline
4 & 2 & 8 & 2 & 1.2 & 0.23\\
\hline
5 & 1 & 6 & 2 & 0.6 & 0.1\\
\hline
6 & 1 & 2 & 2 & N/A & N/A\\
\hline
\end{tabular}
\begin{flushleft}
\end{flushleft}
\label{tab:modelsum}
\end{table}

\paragraph{Model (2)}
\begin{align}
\frac{dP}{dt}&=	\,	k_{pg}P \left(1-\frac{P}{P_\infty}\right)-\frac{k_{pm} s_m P}{\mu_m+k_{mp} P}-k_{pn}f(N)P\\
\frac{dN}{dt}&=\,	\frac{s_{nr}R}{\mu_{nr}+R}-\mu_nN\\
\frac{dD}{dt}&=\,		k_{dn}f_s\left(f(N)\right)-\mu_dD\\
\label{PND_eq}
\end{align}

\paragraph{Model (3)}
 \begin{align}
\frac{dP}{dt}&=	\,	k_{pg}P \left(1-\frac{P}{P_\infty}\right)-\frac{k_{pm} s_m P}{\mu_m+k_{mp} P}-k_{pn}f(N)P\\
\frac{dN}{dt}&=\,	\frac{s_{nr}R_3}{\mu_{nr}+R_3}-\mu_nN,\\
\label{PN_eq}
\end{align}
\quad \quad where 
\begin{align*}
R_3&=f(k_{nn} N+k_{np}P).
\end{align*}

\paragraph{Model (4)}
 \begin{align}
\frac{dN}{dt}&=\,	\frac{s_{nr}R_4}{\mu_{nr}+R_4}-\mu_nN\\
\frac{dD}{dt}&=\,		k_{dn}f_s\left(f(N)\right)-\mu_dD,\\
\label{day_eq}
\end{align}
\quad \quad where 
\begin{align*}
R_4&=f(k_{nn} N+k_{nd}D).
\end{align*}

\paragraph{Model (5)}
\begin{equation}
\frac{dP}{dt}=	k_{pg}P \left(1-\frac{P}{P_\infty}\right)-\frac{k_{pm} s_m P}{\mu_m+k_{mp} P}
\label{eq:mod5}
\end{equation}

\begin{figure}[!ht]
\begin{center}
\includegraphics[width=5.25in]{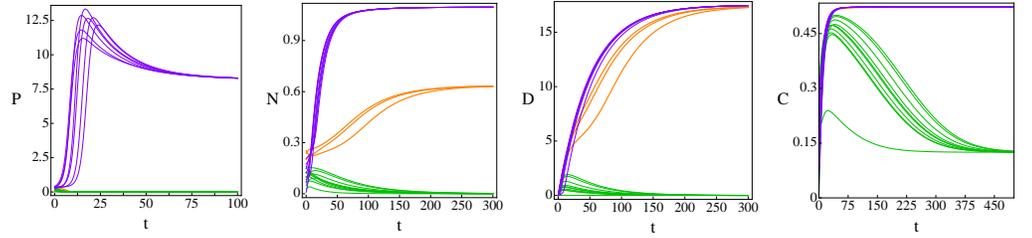}
\end{center}
\vspace{-0.1in}
\caption{{\bf $P$, $N$, $D$, and $C$ versus time for model (1) from
$20$ random initial conditions with no patient variation.} For the
parameter values in Table~\ref{tab:parameters}, model (1) possesses
three fixed points: health (green lines), septic death (purple lines),
and aseptic death (orange lines). The initial conditions are sampled
randomly within the cube: $0\leq P_0\leq 0.42 $, $0\leq N_0\leq
0.255$, $D_0 = 0$, and $0\leq C_0\leq 0.35$. The three fixed points
can be differentiated by the steady-state values of $P$ and $D$:
health ($P=0$, $D=0$), aseptic death ($P=0$, $D>0$), and septic death
($P>0$, $D>0$).  $10$, $7$, and $3$ of the initial conditions evolve to
the health, septic death, and aseptic death fixed points,
respectively.}
\label{fi:initial_conditions}
\end{figure}

In Fig.~\ref{fi:initial_conditions}, we show the time evolution of the
four variables $P$, $N$, $D$, and $C$ for model (1) for twenty
different sets of random initial conditions to illustrate its three
stable fixed points (health, septic death, and aseptic death) using
the parameter values in Table~\ref{tab:parameters}. For trajectories
that approach the septic death fixed point, the pathogen and
neutrophil levels grow rapidly. The high neutrophil level causes
cortisol to increase as well. Despite the high level, the neutrophils
cannot reduce the pathogen load and the cortisol level is not large
enough to reduce the neutrophil level. As a result, the high
neutrophil level causes significant damage at long times, which is
termed septic death due to the associated high pathogen level. Thus,
the septic death steady-state outcome is characterized by $P >0$,
$N>0$, $D>0$, and $C>C_{\infty}$.

In the healthy state, the pathogen level can be reduced to zero by the
neutrophils, and the neutrophil level can be reduced to zero by
cortisol. Once the neutrophil level is zero, the cortisol level
returns to its background level and damage decreases to zero. Thus, the 
healthy state is characterized by $P=0$, $N=0$, $D=0$, and $C=C_{\infty}$.  

During the approach to the aseptic death fixed point, the neutrophil
level is strong enough to reduce the pathogen level to zero, but the
cortisol level is insufficient to reduce the neutrophil level to zero,
which leads to increasing damage. Thus, the aseptic death fixed point is
characterized by $P=0$, $N>0$, $D>0$, and $C>C_{\infty}$.

\begin{figure}[!ht]
\begin{center}
\includegraphics[width=5.25in]{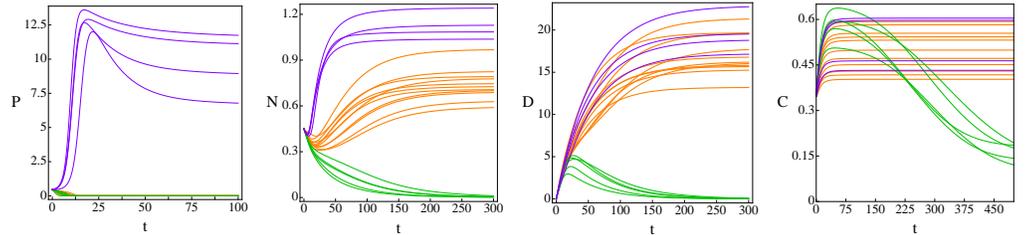}
\end{center}
\vspace{-0.1in}
\caption{{\bf $P$, $N$, $D$, and $C$ versus time for model (1) with
$20$ sets of randomly selected parameters with the same initial
conditions.} $10$\% patient variation allows the system to 
reach the health and septic death fixed points with the initial condition
$P_0=0.45$, $N_0=0.45$, $D_0 = 0$, and $C_0=0.35$, whereas only the aseptic
death fixed point is obtained for this initial condition with no
patient variation. $5$, $4$, and $11$ of the trajectories evolve to
the health, septic death, and aseptic death fixed points,
respectively.}
\label{fi:patient}
\end{figure}

The outcomes of the immune response to infection can vary from patient
to patient, even with the same initial conditions ({\it e.g.}  the
pathogen load). To introduce patient variability into the ODE models,
we select the parameters ($\{q\}$) in models (1)-(5) randomly from
independent Gaussian distributions with mean values $\mu_q$ in
Table~\ref{tab:parameters} and variance $v$ relative to the
mean. Negative values of the parameters can cause the ODE models to
become non-integrable, and thus the parameter distributions are cut
off so that the parameter values are non-negative.  We solve the ODE
models for $10^4$ sets of parameters for each of the $10^4$ random
initial conditions at each $v$. The limits for the sampling of the
initial conditions for each model are given in
Table~\ref{tab:initsum}. We then perform a classification analysis on
these trajectories to predict the steady-state outcomes. The
prediction accuracy $A$ is defined as the number of correct
classifications of the steady-state outcomes divided by the total
number of classifications.

\begin{table}[!ht]
\caption{
{\bf Summary of ODE initial condition ranges}}
\begin{tabular}{|r|l|}
\hline
Model & initial condition ranges \\
\hline
1 &$0\leq P_0 \leq 0.9,\,0\leq N_0 \leq 0.33,\,D_0 =0,\,0\leq C_0 \leq 0.5$\\
\hline
2 & $0 \leq P_0 \leq 0.1,\,0 \leq N_0 \leq 0.15,\,0 \leq D_0 \leq 0.1$\\
\hline
3 &$0 \leq P_0 \leq 0.2,\,0 \leq N_0 \leq 0.3$ \\
\hline
4 & $0 \leq N_0 \leq 0.15,\,0 \leq D_0 \leq 0.1$\\
\hline
5 &$0 \leq P_0 \leq 0.7$ \\
\hline
6 & $0 \leq x_0 \leq 2\pi$\\
\hline
\end{tabular}
\begin{flushleft}
\end{flushleft}
\label{tab:initsum}
\end{table}
\section*{Results} 

For a deterministic system of ODEs, the basin of attraction for a given
fixed point is defined as the collection of initial conditions that
evolve to that particular fixed point.  For a given set of parameters,
each of the ODE models (1)-(5) possesses well-defined
(non-overlapping) basins of attraction for each fixed point.

However, different outcomes can be achieved even for a single initial
condition if the parameters of the ODE model are varied. (See
Fig.~\ref{fi:patient}.)  For example, the ratio of the parameters
$s_c$ and $\mu_c$ determines the background level of cortisol in model
(1). Background cortisol levels are known to vary from patient to
patient and can vary from one organ system to another in a given
patient. To mimic these variations, we select sets of parameters
randomly with mean values in Table~\ref{tab:parameters} and variances
$v$ relative to their mean values. (See Materials and Methods.) With
patient variation, an initial condition can possess multiple outcomes,
and thus the basins of attraction for the fixed points overlap as
shown in Fig.~\ref{fi:schematic}.

\begin{figure}[!ht]
\begin{center}
\includegraphics[width=5.25in]{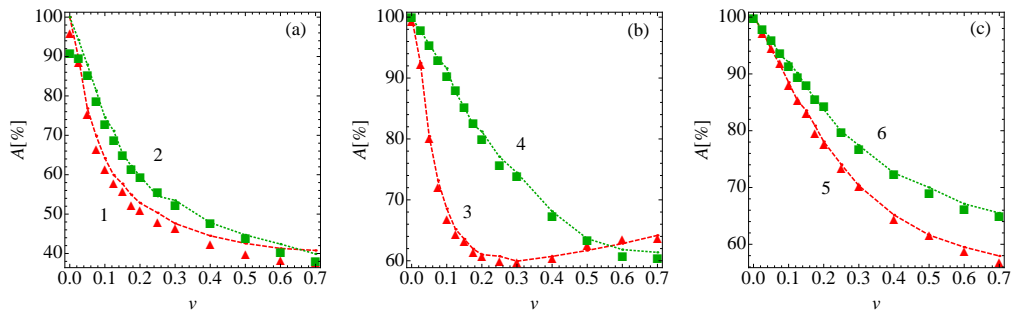}
\end{center}
\caption{ {\bf Prediction accuracy of the steady-state outcome as a
function of patient variation.} The prediction accuracy $A$ using a
logistic regression classifier at time $t_c=0$ (symbols) and the average
best guess over $10^3$ initial conditions (dashed curves) versus
patient variation $v$ for (a) models (1) and (2), (b) models (3) and
(4), and (c) models (5) and (6).}
\label{fi:accvsvar}
\end{figure}

We seek to predict the patient steady-state outcomes in models (1)-(5) in the
presence of patient variability $v$. For the prediction method, we
employ logistic regression with one-versus-all
classification~\cite{Hastie2009}.  We compare the prediction accuracy
at patient variability $v$ to the average best guess of the steady-state
outcome.  For the best guess method, we determine the steady-state
outcome for each of $10^2$ sets of parameters for a given initial
condition. We define the best guess as the steady-state outcome with
the highest number of occurrences and record the frequency $f_i$ of the
best guess for initial condition $i$.  We then average the frequency
$f_i$ over $10^3$ initial conditions for each $v$ to obtain an
estimate for the prediction accuracy in systems with basin overlap.

For the prediction method, we solve a given system of ODEs for
$N_i=10^4$ random initial conditions, each with randomly selected
parameter sets with variance $v$. We choose $N_t=800$ of the $N_i$
trajectories randomly to train the classifier and predict the outcome
of the remaining $9200$ trajectories. The classifier maps the state of
the system at a given time $t_c$ to a particular steady-state
outcome. The prediction accuracy is then averaged over $10$ training
and prediction runs, each with $N_t=800$ randomly selected training
trajectories.

\begin{figure}[!ht]
\begin{center}
\includegraphics[width=5.25in]{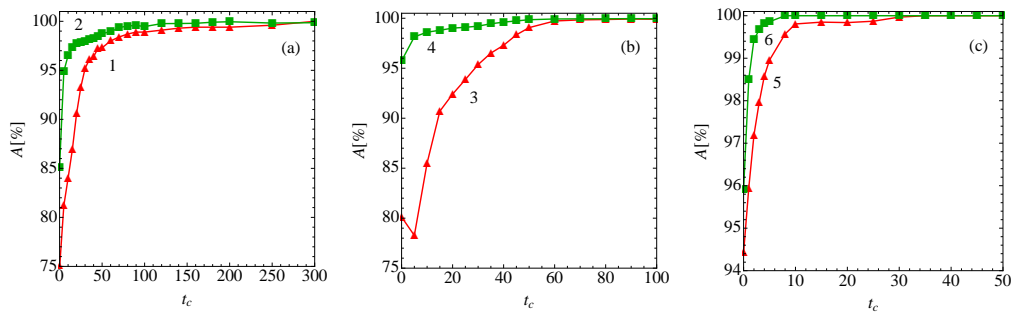}
\end{center}
\caption{ {\bf Prediction accuracy of the steady-state outcome as 
a function of the classification time.} The prediction accuracy $A$ using a
logistic regression classifier at time $t_c$ (symbols) for (a) models (1) and 
(2), (b) models (3) and (4), and (c) models (5) and (6) for patient 
variation $v=0.05$.}
\label{fi:accvsT}
\end{figure}

In Fig.~\ref{fi:accvsvar}, we compare the accuracy $A$ of the logistic
regression prediction method (with classification at time $t_c=0$) to
the average best-guess frequency as a function of the patient
variability $v$ for models (1)-(5). For all model ODEs, the prediction
accuracy for the logistic regression prediction method is near $100\%$
at $v=0$, decreases for increasing patient variability, and reaches a
plateau near $1/n_f$ in the large $v$ limit, where $n_f$ is the number
of stable fixed points in the model (except for model (3)). For model (3)
with two steady-state outcomes, the prediction accuracy is
non-monotonic and increases for $v>0.3$ because this ODE
model begins to sample parameter regimes where one steady-state
outcome is much more probable than the other. In addition, for all
models the average best-guess frequency provides an upper bound for
the accuracy of the prediction algorithm.  Hence, the overlap of the
basins of attraction imposes a limit on the prediction accuracy.
To test the generality of these results, we studied another  
one-dimensional ODE (model (6)) with varied fixed point
structure compared to that for model (5). (See Supporting Information.)
The results for model (6) are very similar to those for models (1)-(5).

\begin{figure}[!ht]
\begin{center}
\includegraphics[width=5.25in]{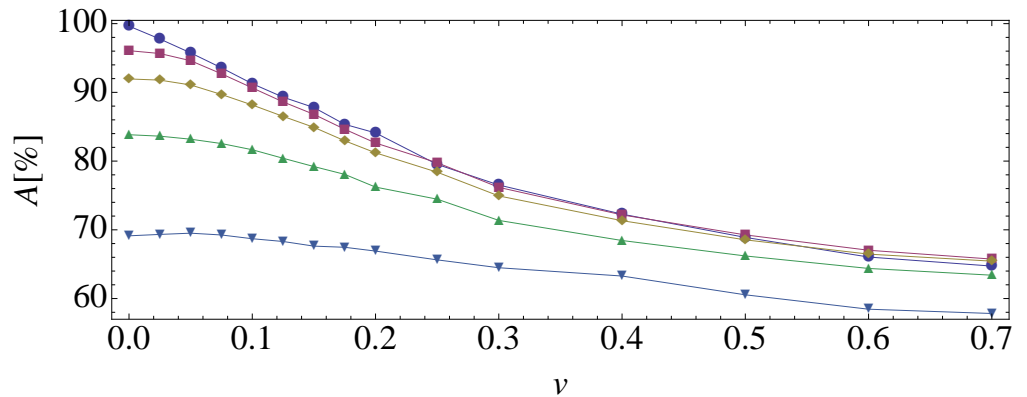}
\end{center}
\caption{{\bf Prediction accuracy as a function of patient variation
for different noise strengths.} The prediction accuracy $A$ using a
logistic regression classifier at time $t_c=0$ for model (6) in the 
presence of measurement noise with strength $s=0$ (circles), $0.05$ (squares),
$0.10$ (diamonds), $0.20$ (triangles), and $0.50$ (triangles).}
\label{fi:model6_noise}
\end{figure}

In Fig.~\ref{fi:accvsvar}, we showed results for the logistic
regression prediction method with classification at $t_c=0$. In
Fig.~\ref{fi:accvsT}, we show the prediction accuracy for models
(1)-(6) with patient variability $v=0.05$ as a function of the
classification time $t_c$. For all models, the prediction accuracy
grows with increasing $t_c$, reaching nearly $100\%$ beyond a
characteristic time $t^*$ that depends on the model.  The prediction
accuracy improves at later classification times because the system
trajectories have moved closer to the fixed points and hence the
basins of attraction are more easily separated as shown in
Fig.~\ref{fi:schematic} for model (1).

We also investigated the variation of the prediction accuracy in the
presence of measurement noise. We took the trajectories generated for
Fig.~\ref{fi:accvsvar} and added Gaussian random noise to the model
variables with variance $s$ at each time point. We then performed
training and testing on the noisy data with classification at time
$t_c=0$. In Fig.~\ref{fi:model6_noise}, we show for model (6) that the
prediction accuracy decreases with increasing $s$. We find similar
results for models (1)-(5). These results emphasize that even if the
measurement noise could be reduced to zero, the patient variation
imposes an intrinsic limitation to outcome prediction.

\section*{Discussion}

In clinical settings it is of great importance to determine patient
outcomes as quickly as possible with maximum accuracy. In this
manuscript, we studied the effects of patient variability on the
ability to predict steady-state outcomes in systems of ODEs that model
the immune response to infection.  For deterministic systems of ODEs
with a given fixed set of parameters, each initial condition can be
mapped to a given steady-state outcome (or fixed point) and the
collection of initial conditions that map to a given steady-state
outcome is defined as the basin of attraction of that outcome. Each
virtual patient can be defined by a given set of parameters in the
model ODE and patient variability can be introduced by varying the
model parameters.

We showed that the introduction of patient variation leads to overlaps
of the basins of attraction for the steady-state outcomes.  In
particular, a given initial condition can map to multiple steady-state
outcomes for different virtual patients ({\it i.e.} $v>0$), which is
similar to the case of patients showing different responses to
infection in clinical settings. We find that the prediction accuracy
of the outcomes decreases strongly with increasing patient
variability.  Our results emphasize that even when the complete state
of the system is known ({\it i.e.} all patient variables are measured
precisely as a function of time), we have limited knowledge of the
patient outcome when there is patient-to-patient variability that
gives rise to basin overlap.

Our results also show that for all of the model ODEs studied the
prediction accuracy increases as the time $t_c$ used for
classification increases. As $t_c$ increases, the systems move closer
to their steady-state outcomes and the basins of attraction separate,
which increases the prediction accuracy.  Again, this result is
consistent with clinical experience. If a clinician waits to see if
the condition of the patient improves or worsens, the prognosis will
become more accurate. In our work, we explicitly show that
patient-to-patient fluctuations cause a competition between early and
accurate outcome prediction.

In this work, we focused on discrete steady-state outcomes ({\it i.e.}
health or death of the patient) of the immune response to
infection. However, in many biomedical scenarios, the outcomes involve
continuous variables rather than discrete states. In future work, we
will apply similar techniques to understand the effects of patient
variability on the predictions of continuous model variables, for
example, the immune response and vaccination efficacy for
influenza~\cite{hancioglu}.

\section*{Supporting Information}

In this section, we describe a generalized one-dimensional ODE that
has the same fixed point structure as model (5) (Eq.~\ref{eq:mod5}),
but different locations for the fixed points. We also provide
technical details for the logistic regression one-versus-all
classifier employed in this work.

\begin{figure}[!ht]
\begin{center}
\includegraphics[width=5.25in]{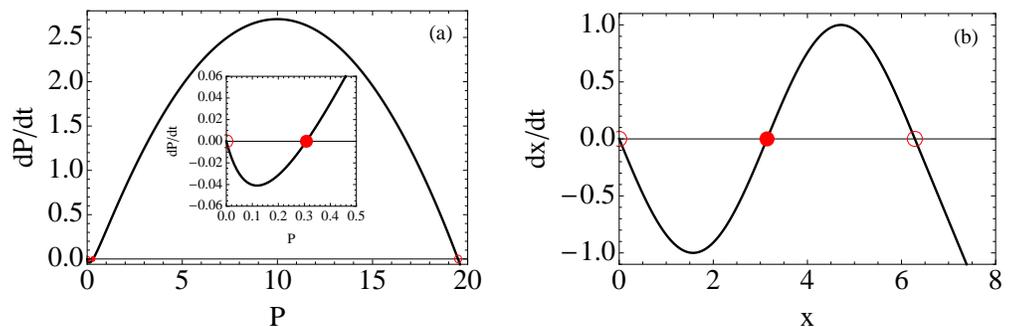}
\end{center}
\caption{ {\bf Comparison of ODE models (5) and (6).} The functions
$f(P)=dP/dt$ and $f(x)=dx/dt$ for models (a) (5) and (b) (6). (See
Eqs.~\ref{eq:mod5} and~\ref{eq:mod6}.) Stable and unstable fixed
points are marked by open and filled circles, respectively. Both
models share the same fixed point topology: one stable fixed point at
zero and one at a positive value. The unstable fixed
point lies between the two stable fixed points. The inset in (a) magnifies
$dP/dt$ near the origin.}
\label{fi:mod5_mod6}
\end{figure}

Model (6) is a one-dimensional ODE for the variable $x$:
\begin{equation}
dx/dt = 
\begin{cases}B\, {\rm Cos}(k\,x +\pi/2),& \text{if } 0\leq x \leq 2\pi/k\\
- B\,k\,{\rm Sin}(5\pi/2) \left(x-2\pi/k\right),& \text{if }2\pi/k < x
\end{cases}
\label{eq:mod6}
\end{equation}
with two parameters $k=1$ and $B=1$. Model (5), which is a
one-dimensional ODE for pathogen $P$, possesses three fixed points:
$P=0$, $0.3078$, and $19.49$ for the mean parameters in
Table~\ref{tab:parameters}.  As shown in Fig.~\ref{fi:mod5_mod6} (a),
the two outer fixed points for model (5) are stable, and the middle
fixed point, which is near zero, is unstable.  For model (6), the
central unstable fixed point is moved to the midpoint of the two outer
stable fixed points and the shape of the function is changed to retain
the fixed point structure.

For the prediction method, we train a logistic regression classifier
in a one-versus-all scheme to predict the steady-state outcomes based
on the values of the set of variables
$x^{(i)}=\{x^i_1,x^i_2,\dots,x^i_N\}$ at a specified time $t_c$, where
$N$ is the number of variables in the model. To distinguish between
two outcomes with logistic regression, we label the outcomes,
$y^{(i)}=0$ and $y^{(i)}=1$ for the $i$th set of variables. We
fit a logistic (or sigmoidal) function to the input data such that the
function
\begin{equation}
\label{one}
P_{\theta}(x^{(i)})=\frac{1}{1+e^{\theta_0+\theta_1\,x^i_1+\theta_2\,x^i_2+\dots+\theta_N\,x^i_N}}
\end{equation}
gives the probability that $y^{(i)}=1$ given the input data
$x^{(i)}$. The parameters $\theta_j$, where $j=1,2,\ldots,N$,
are determined by minimizing the cost function
\begin{equation}
\label{two}
J(\theta_1,\theta_2,\ldots,\theta_N)=-\frac{1}{m}\left[\sum_{i=1}^my^{(i)}{\rm log}(P_{\theta}(x^{(i)}))+(1-y^{(i)}){\rm log}(1-P_{\theta}(x^{(i)}))\right],
\end{equation}
where $m$ is the number of training samples. Since models (1) and (2)
possess three steady-state outcomes (aseptic death, septic death, and
health), we must go beyond the binary classification scheme described
above. To classify ODE models with three steady-state outcomes, we
implement the one-versus-all classification scheme. To do this, we
consider three labeled outcomes, $y^{(i)}_h$, $y^{(i)}_{ad}$, and
$y^{(i)}_{sd}$, for a given set of variables $x^{(i)}$. $y^{(i)}_h =
0$ if the patient outcome is not health ({\it i.e.} aseptic or septic
death) and $y^{(i)}_h = 1$ if the patient outcome is health. Similar
definitions apply for $y^{(i)}_{ad}$ and $y^{(i)}_{sd}.$ (See
Table~\ref{tab:onevsall}.) We use these labeled outcomes and
Eq.~\ref{two} to determine $P_{h}(x)$, $P_{ad}(x)$, and $P_{sd}(x)$.
Given an unlabeled set of variables ($x=\{x_1,x_2,\ldots,x_N\}$),
we calculate $P_h$, $P_{ad}$, and $P_{sd}$ and select the outcome with
highest probability to be the predicted outcome.

\begin{table}[!ht]
\caption{
{\bf One-versus-all outcomes}}
\begin{tabular}{|r|l|}
\hline
Class 1 ($y=0$) & Class 2 ($y=1$)\\
\hline
septic death + health & aseptic death\\
\hline
health + aseptic death & septic death\\
\hline
 aseptic death + septic death & health\\
 \hline
\end{tabular}
\label{tab:onevsall}
\end{table}

\section*{Acknowledgments}
This work was partially supported by DARPA (Space and Naval Warfare
System Center Pacific) under award number N66001-11-1-4184 and
partially supported by the National Science Foundation Grant No.
NSF PHY11-25915. Additional support for this work was provided by the
Infectious Disease Supercluster, Colorado State University, 2012 seed
grant. This work also benefited from the facilities and staff of the
Yale University Faculty of Arts and Sciences High Performance
Computing Center and NSF Grant No. CNS-0821132 that partially funded
acquisition of the computational facilities.

\nolinenumbers

%
%
%

\end{document}